\documentclass[twocolumn,showpacs,superscriptaddress]{revtex4}  % for review and submission 

\usepackage{graphicx}  % needed for figures
\usepackage{dcolumn}   % needed for some tables
\usepackage{bm}        % for math
\usepackage{amssymb}   % for math

\usepackage{verbatim}

\begin{document}

\title{All-electrical injection and detection of a spin polarized current using 1D conductors}

\author{T.-M. Chen}
\email{tmchen@mail.ncku.edu.tw}
\affiliation{Department of Physics, National Cheng Kung University, Tainan 701, Taiwan}
\affiliation{Cavendish Laboratory, J J Thomson Avenue, Cambridge CB3 0HE, United Kingdom}
\author{M. Pepper}
\affiliation{Department of Electronic and Electrical Engineering, University College London, London, WC1E 7JE, United Kingdom}
\author{I. Farrer}
\author{G. A. C. Jones}
\author{D. A. Ritchie}
\affiliation{Cavendish Laboratory, J J Thomson Avenue, Cambridge CB3 0HE, United Kingdom}

\begin{abstract}
All-electrical control of spin transport in nanostructures has been the central interest and challenge of spin physics and spintronics. Here we demonstrate on-chip spin polarizing/filtering actions by driving the gate-defined one dimensional (1D) conductor, one of the simplest geometries for integrated quantum devices, away from the conventional Ohmic regime. Direct measurement of the spin polarization of the emitted current was performed when the momentum degeneracy was lifted, wherein both the 1D polarizer for spin injection and the analyzer for spin detection were demonstrated. The results showed that a configuration of gates and applied voltages can give rise to a tunable spin polarization, which has implications for the development of spintronic devices and future quantum information processing.
\end{abstract}

\maketitle
%%%%%%%%%%%%%%%%%%%%%%%%%%%%%%%%%%%%%%%%%%%%%%%%%%%%%%%%%%%%%%%%%%%%%
%% Start the main part of the manuscript here.
%%%%%%%%%%%%%%%%%%%%%%%%%%%%%%%%%%%%%%%%%%%%%%%%%%%%%%%%%%%%%%%%%%%%%

\textbf{}

There is considerable interest in being able to control spin dynamics, particularly in mesoscopic and nanoscale semiconductor devices\cite{zutic_rmp04,fert_rmp08} as this could lead to the development of a range of electronic functions not presently available. In order to develop successfully such concepts it is necessary to controllably generate, manipulate, and detect spin currents by electrical means and so minimize, or eliminate, the use of ferromagnetic contacts or external magnetic fields. Most research towards the implementation of this electrical approach has focussed on using the spin-orbit interaction to induce spin polarized transport, as reported in various nanostructures\cite{kato_science04,konig_science07,hsieh_nature08,frolov_nature09} including one-dimensional (1D) conductors\cite{debray_naturenano09,qua_naturephys10}. However, it is essential to develop a more general approach in which materials with a strong intrinsic spin-orbit coupling are no longer necessary, and consequently a longer spin dephasing (relaxation) time will be obtained of crucial importance for quantum information processing.

In theory, it is possible to produce transition from anti-ferromagnetic to ferromagnetic behaviour by controlling the exchange interaction, although this can be difficult to achieve in practice. If such a mechanism could be successfully utilized for on-chip spin injection, the problems associated with conventional methods of spin injection --- such as the impedance mismatch, which drastically limits the spin polarization (spin pumping efficiency) of the injected current\cite{schmidt_prb00} --- can be avoided. Furthermore, the fast-gating technique, which has been well developed in conventional microelectronics, allows it to be used for rapid control of the spin content.

Studies of quasi one-dimensional conduction\cite{thornton_prl86} have been of interest for a considerable time due to its strong electron-electron interaction, much of this work has been with reference to the spin properties\cite{thomas_prl96,micolich_jpcm11}. The variation of the current with the dc source-drain voltage has been shown to be particularly useful in providing quantitative measurements on the energies of the 1D subbands in the channel. For ballistic transport this voltage is dropped at the two ends of the channel and lifts the momentum degeneracy, and has been used, for example, to derive the value of the Lande \textit{g} factor by measuring the spin splitting in a magnetic field\cite{patel_prb91a,chen_nanolett10}. It has also been used to show that there is a spontaneous lifting of the spin degeneracy in the absence of a magnetic field, which is related to the 0.7 structure\cite{chen_nanolett10}.

Furthermore, there is a feature which appears as a plateau, or structure, with increasing dc source-drain voltage at, or near, the value of $0.25(2e^2/h)$ in the differential conductance. Although this feature was apparent in early work on one-dimensionality\cite{patel_prb91b,thomas_philmag98,kristensen_prb00}, it was in general regarded as a spin degenerate state with a decreased differential conductance\cite{picciotto_prl04,kothari_jap08,ihnatsenka_prb09}. However, it was recently proposed that the $0.25(2e^2/h)$ feature could be a consequence of a lifting of both momentum and spin degeneracy\cite{chen_apl08}. The loss of the momentum degeneracy on its own producing a value of $e^2/h$ and an absence of spin degeneracy accounting for the remaining factor of $1/2$. This is a very surprising result of increasing the source-drain voltage, and in order to substantiate this conclusion it is crucial to provide direct evidence of spin polarization which does not rely on an inference from conductance plateaux, particularly because it has been suggested that it is possible for the differential conductance value to be reduced in the non-Ohmic regime\cite{picciotto_prl04,kothari_jap08,ihnatsenka_prb09}.

\begin{figure}[t!]
\begin{center}
\includegraphics[width=0.95\columnwidth]{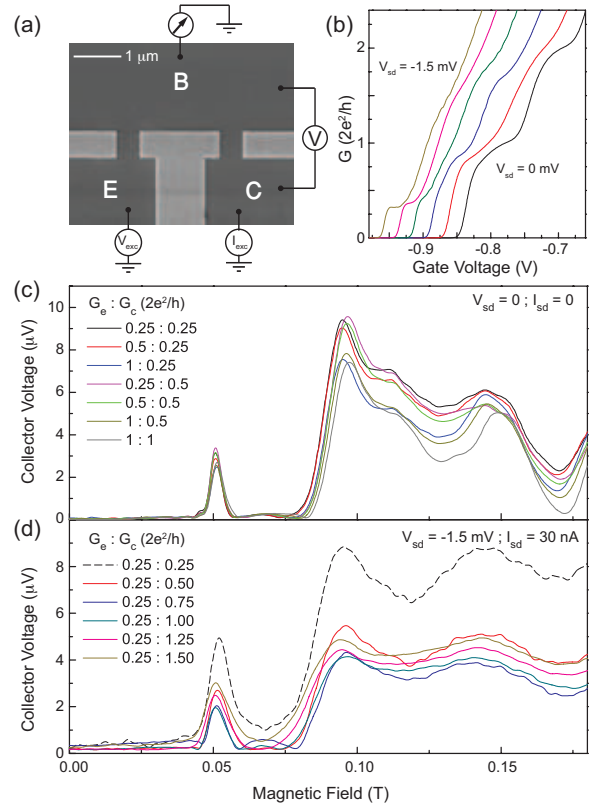}
\end{center}
\caption[Figure 1]
{\small (a) Micrograph and electric circuit showing the emitter-collector configuration used in the experiment. Electrons are focused by a small perpendicular magnetic field and travel from the emitter (E), through the 2D base region (B) into the collector (C). (b) The differential conductance of the emitter at $B = 0$ with application of a source-drain bias from $V_{sd}$ = 0 to $-1.5$~mV in steps of $-0.3$~mV. The conductance anomaly at around $G = 0.25(2e^2/h)$ starts when $V_{sd} = -0.9$~mV. Data are offset for clarity. (c) The collector voltage (focusing signal) for $V_{sd} = I_{sd} = 0$; the focusing voltage is nearly independent of the conductance of both emitter and collector from $2e^2/h$ to $0.25(2e^2/h)$. (d) A substantial rise in the focusing signal appears when both emitter and collector were set to $G = 0.25(2e^2/h)$ and the dc source-drain biases $V_{sd} = -1.5$~mV and $I_{sd}=30$~nA were applied respectively.
} 
\label{fig1}
\end{figure}

In this work, we have utilized a technique of electron focusing\cite{potok_prl02,folk_science03,rokhinson_prl04,rokhinson_prl06} to directly measure the degree of spin polarization of the current. The focusing device geometry is shown in Figure~\ref{fig1}(a), wherein a small perpendicular magnetic field $B_{\perp}$ is applied to bend and inject ballistic electrons from an emitter, a short one dimensional region formed by split gates (quantum point contact), which acts as a spin polarizer in this work. The electrons pass through the two-dimensional base region, which is grounded, into the collector which is an identical device to the emitter; in the context of this experiment the collector acts as a spin analyzer. With current flowing into the device from the emitter, and with the base connected to ground, the collector-base voltage shows periodic peaks as a function of $B_{\perp}$ which is due to the focusing of electrons into the collector. These focusing peaks occur whenever an integer multiple of the cyclotron diameter, $2m^*v_F/eB_{\perp}$, where $m^*$ is the electron effective mass and $v_F$ is the Fermi velocity, equals the distance, $L$, between emitter and collector.

As the collector is not connected to ground, a voltage $V_c = I_c / G_c$ develops between the collector and base, where $I_c$ is the current flowing through the collector which has conductance $G_c$. Both the conductance and current can be further written as $G_c=e^2/h (T_\downarrow + T_\uparrow)$ and $I_c = \alpha I_e (T_\downarrow + T_\uparrow)$ where the arrows represent the electron spins, $I_e = I_\downarrow + I_\uparrow$ is the current injected from the emitter, $T_\downarrow$ ($T_\uparrow$) is down-spin (up-spin) transmission of the collector and $\alpha$ is a parameter, accounting for spin-independent imperfections during the focusing process\cite{potok_prl02}.

This situation has been considered by Potok \textit{et al.}\cite{potok_prl02}, who have shown that a simple derivation gives the magnitude of the height of the peaks in collector-base voltage. This can be written in terms of the degree of spin polarization induced by the emitter $P_e = (I_\downarrow - I_\uparrow)/(I_\downarrow + I_\uparrow)$ and the spin selectivity of the collector $P_c = (T_\downarrow - T_\uparrow)/(T_\downarrow + T_\uparrow)$. They found the following relation 
\vspace*{3mm}\begin{equation}
V_c = \alpha \frac{h}{2e^2} I_e (1 + P_e P_c),
\label{eq1}
\vspace*{3mm}\end{equation}
which was confirmed by inducing a Zeeman spin splitting with a strong in-plane magnetic field\cite{potok_prl02,folk_science03}. Consequently, if both emitter and collector are spin polarized the collector voltage is doubled compared to when either emitter or collector allows spin degeneracy.

Here we investigated the spin balance in the focusing stream as the conductances of both emitter and collector were varied in the absence of a magnetic field (except for the small focusing field $B_{\perp}$). The particular objective was to clarify the spin content of the current when the differential conductance was in the region of the 0.25 plateau. This work used samples comprising a high-mobility two-dimensional electron gas formed at the interface of GaAs/Al$_{0.33}$Ga$_{0.67}$As heterostructures. The low temperature mobility was $2.3 \times 10^6$~cm$^2$/Vs at a carrier density $1.17 \times 10^{11}$~cm$^{−2}$ giving a mean free path for momentum relaxation $\sim 13$~ $\mu$m. This is much longer than the focusing path, although we note that the small angle scattering length is much less and may contribute to a broadening of the focusing peak.

Measurements were performed at a temperature of 80~mK, the electrical connections are shown in Figure~\ref{fig1}(a). Two devices were measured and gave similar and reproducible results. Simultaneous lock-in measurements of the emitter and collector conductances and the focusing signal were performed by applying two independent excitation sources of (i) a 77 Hz ac voltage 20~$\mu$V with a dc bias $V_{sd}$ applied to the emitter and (ii) 31 Hz ac current 1~nA with a dc bias $I_{sd}$ applied to the collector. It was verified that the focusing signal $V_c$ was linear with current $I_e$; for clarity, all the data presented here was rescaled for $I_e = 1$~nA. The current-bias excitation, i.e., source (ii), is required to prevent the collector from sinking injected current, as well as increasing the bias across the collector pushing it into the 0.25 regime.

%insert Fig.1 %
%continued%

Both the emitter and collector show one-dimensional conductance quantisation and a source-drain voltage induced plateau at $0.25(2e^2/h)$ at $B = 0$\cite{chen_apl08}, as shown in Figure~\ref{fig1}(b). The measured focusing peaks are shown in Figure~\ref{fig1}(c) when the emitter and collector are set at the described values of conductance for $V_{sd} = I_{sd} = 0$. Focusing peaks appear periodically, at intervals of $B_\perp = 0.05$~T, which is consistent with the cyclotron motion $B_{\perp}=2m^*v_F/eL$ calculated from the two-dimensional electron concentration. The height of the focusing peaks, as anticipated, barely changes with decreasing conductance of both emitter and collector from $2e^2/h$ to $0.25(2e^2/h)$, indicating that there is no change in their spin polarization, i.e., $P_e$ and/or $P_c = 0$. As expected the peak height is independent of $G_e$ and $G_c$ for constant current $I_e$ injected from the emitter point contact.

When a dc bias is applied across the emitter and collector the focusing peak exhibits very different behaviour to that previously observed at zero bias. In Figure~\ref{fig1}(d), the focusing peaks are shown for various $G_c$ when $G_e$ is fixed at $0.25(2e^2/h)$, and when the dc biases across the emitter and collector were set at $V_{sd} =-1.5$~mV and $I_{sd} = 30$~nA, respectively. It was observed that the peak height barely changes with decreasing $G_c$ from $1.5(2e^2/h)$ to $0.5(2e^2/h)$, but rises considerably when this approaches $0.25(2e^2/h)$, i.e., where the anomalous plateau is found as shown in Figure~\ref{fig1}(b). This substantial rise is predicted by Equation~(1), if there is an increasing degree of spin polarization in both the emitter and collector.

%insert Fig.2 %
\begin{figure}[t!]
\begin{center}
\includegraphics[width=0.95\columnwidth]{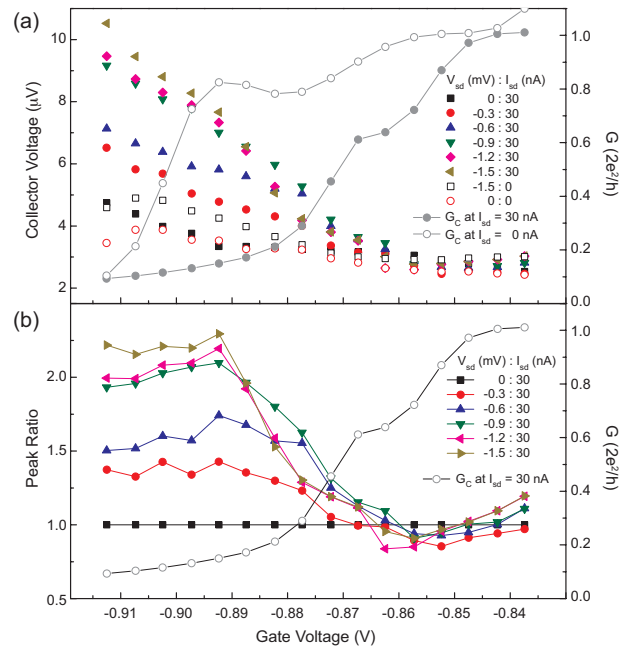}
\end{center}
\caption[Figure 2]
{\small (a) Voltage of the first focusing peak, left, and the corresponding conductance of the collector, right, as a function of gate voltage, for various values of dc source-drain bias $V_{sd}$ and $I_{sd}$. Enhancement of focusing peaks only occurs near the 0.25 conductance region when dc source-drain biases are applied across both the emitter and collector. (b) Normalized peak ratio and corresponding conductance as a function of gate voltage, with $I_{sd}$ set to 30~nA and $V_{sd}$ swept from 0 to $-1.5$~mV. The peak ratio rises and then saturates when both conductance and dc source-drain bias are at the appropriate values for the appearance of the $0.25(2e^2/h)$ plateau.
} 
\label{fig2}
\end{figure}
%continued%

Figure~\ref{fig2}(a) shows the height of the first peak as both the dc biases and the collector conductance were varied with the emitter conductance locked at $G_e = 0.25(2e^2h)$; this peak was chosen for investigation because of its robust structure and is seen to stay fairly constant at $\sim 3$~$\mu$V, essentially independent of both $V_{sd}$ and $I_{sd}$, when $G_c \sim 2e^2/h$. However, in the low conductance region when $G_c \leq 0.25(2e^2/h)$, and $I_{sd} = 30$~nA, the focusing peaks increase as the dc bias is increased, negatively, from $V_{sd}=0$ and then saturates when $V_{sd}$ is near $-0.9$~mV. The focusing peaks at $ |V_{sd}| \geq 0.9$~mV are approximately twice the value of those at $V_{sd} = 0$ for every individual value of collector conductance below $\sim 0.25(2e^2/h)$. This, according to Eq.~(1), implies that both emitter and collector are fully spin polarized, i.e., $P_e = P_c = 1$. The saturation of the peak height is also consistent with the fact that both $P_e$ and $P_c$ cannot be larger than 1.

To further verify this bias-induced spin polarization, $I_{sd}$ was decreased from 30~nA to 0 with $V_{sd}$ still at $-1.5$~mV. Figure~\ref{fig2}(a) shows that the height of the focusing peaks drops back to almost the same value obtained when $V_{sd} = 0$ and $I_{sd} = 30$~nA as well as when both $V_{sd}$ and $I_{sd}$ are zero. This is again expected when either polarizer or analyzer are spin degenerate (i.e., either $P_e$ or $P_c$ equals 0). Finally, it is important to note that the value of source-drain bias $V_{sd} =-0.9$~mV at which the focusing peak height $V_c$ saturates is consistent with the bias at which the 0.25 anomaly appears, as shown in Figure~\ref{fig1}(b).

The evolution of the focusing peaks as a function of conductance\cite{note1} is also shown in Figure~\ref{fig2}(a), the focusing peak rises as $G_c$ is reduced below $2e^2/h$, but the manner of the increase varies for different dc source-drain biases. At $V_{sd} = 0$ and $I_{sd} = 30$~nA, the peak voltage barely increases until $G_c$ is reduced below $\sim 0.15(2e^2/h)$, in the near-pinch-off region, whereas at $V_{sd} < 0$ and $I_{sd} = 30$~nA the peak voltage starts to increase once $G_c$ is reduced below $\sim 0.6(2e^2/h)$. The near-pinch-off increase in peak voltage with the reduced value of $G_c$ could be attributed to an $\alpha$-dependent enhancement; this has been suggested previously when $G_c$ is low\cite{potok_prl02} although the origin is not clear.

To remove nonspin related effects from the focusing peak, all the peak voltages are normalized by the values at $V_{sd} = 0$ and $I_{sd} = 30$~nA\cite{note2}. Figure~\ref{fig2}(b) shows the normalized peak ratio, proportional to $(1+ P_e P_c)$, and the corresponding conductance as a function of gate voltage, with $I_{sd}$ set to 30~nA and $V_{sd}$ swept from 0 to $-1.5$~mV. As seen the peak values rise with reducing conductance and then saturate when $G_c$ reaches the region of the 0.25 plateau, suggesting that $P_c$ has reached its maximum value of 1. Similarly, the peak ratio in the 0.25 regime rises with increasing source-drain bias applied across the emitter and then saturates. This reaches a value of $\sim 2$, at $V_{sd}=-0.9$~mV when the 0.25 feature appears in the emitter conductance.

%insert Fig.3 %
\begin{figure}[t!]
\begin{center}
\includegraphics[width=0.95\columnwidth]{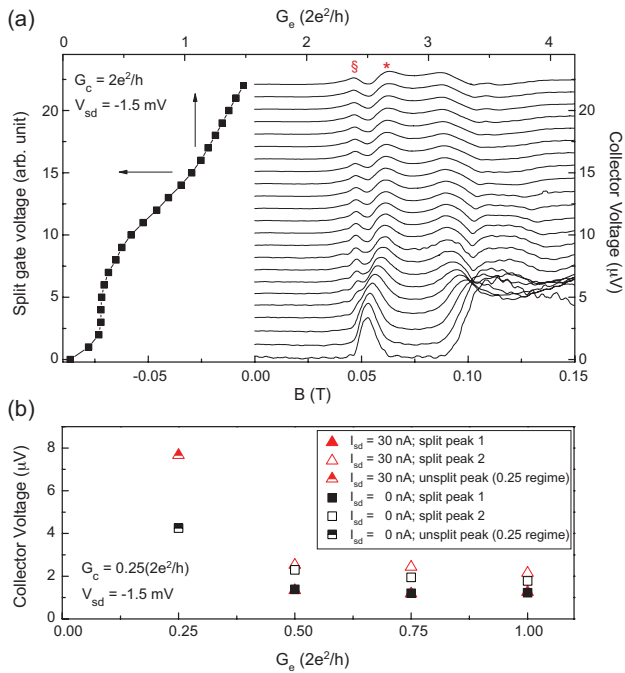}
\end{center}
\caption[Figure 3]
{\small (a) The focusing peak voltage (right half) versus B-field for several emitter conductance $G_e$ from 0 to $1.5(2e^2/h)$, along with its corresponding value of $G_e$ (left half) when the source-drain bias $V_{sd}$ = -1.5~mV is applied across the emitter. The focusing voltage is offset for clarification. The focusing peak starts to split in two when the $G_e$ starts to rise from the $0.25(2e^2/h)$ plateau. (b) Voltage of the first focusing peak versus $G_e$ when $G_c$ is locked at the 0.25 plateau. Both the split peak 1, the peak marked by $\S$ in (a), and the split peak 2, the peak marked by $\star$ in (a), barely change with $I_{sd}$ and $G_e$. Enhancement of focusing peaks only occurs when both $G_e$ and $G_c$ are at the 0.25 regime.
} 
\label{fig3}
\end{figure}
%continued%

Such focusing peak enhancement was further verified by another set of measurement where $G_c$ was locked at $0.25(2e^2/h)$ as both the dc biases and $G_e$ were varied. Figure~3(a) shows that the focusing peak splits when the emitter conductance $G_e$ is above the $0.25(2e^2/h)$ plateau, indicating that both the source and the drain chemical potential have reached a 1D subband. The two split peaks represent focusing electrons with the source and the drain potential, respectively, whereas in the 0.25 and zero-bias regions no splitting occurs because there is only one potential across the 1D subband.

Indeed the peak enhancement, which is the evidence of spin polarization, occurs only when both $G_e$ and $G_c$ are in/below the bias-induced 0.25 plateau. Figure~3(b) clearly shows that the split focusing peaks barely changes with $G_e$ [only when $G_e > 0.25(2e^2/h)$] and $I_{sd}$ when $G_c$ is locked at $0.25(2e^2/h)$. The enhancement only occurs when $G_e = 0.25(2e^2/h)$ and the source-drain bias $I_{sd}$ is applied. In addition, it is noteworthy that the spin-polarized $0.25(2e^2/h)$ plateau is very robust, nearly independent of temperatures up to $4.2$~K.

These results show that the emitter is functioning as a spin polarizer and the collector as a spin analyzer, demonstrating that a manipulation of the degree of polarized spin current can be achieved by tuning the source-drain bias at low values of conductance. For instance, Figure~\ref{fig2}(b) shows that the spin polarization of the injecting current $P_e$ was $\sim30\%$ when the emitter was set to $V_{sd} = -0.3$~mV and $G_e= 0.25(2e^2/h)$, whereas $P_e$ reaches $60\sim70\%$ for $V_{sd} = -0.6$~mV. We note that in the region of the 0.7 anomaly, which is found in the absence of bias, the enhancement of the peak height will be $\sim 10\%$ and difficult to observe unambiguously.

The effects observed here indicate that the non-equilibrium electron energy distribution and the spin coherence are maintained during the focusing transit into the collector. The transit time is sufficiently short (approximately 20 picoseconds) that phonon emission is not occurring to any significant degree, so allowing all the emitted electrons to enter the collector. The spin coherence length exceeds the path length so that the spin polarization is maintained during the focusing which augurs well for applications of this phenomenon.

Our experiments establish a link between spin and momentum which is unusual in the system with weak spin-orbit coupling. It seems most likely that the cause of the 0.25 is that a spin polarized stream of electrons is the lowest energy configuration; this configuration is retained as there is only one direction of momentum and an absence of spin scattering by electrons with the opposite momentum. A physical mechanism based on exchange interaction has recently been proposed for the 0.25 anomaly which explains the lifting of the spin degeneracy\cite{lind_prb11} in the regime of non-equilibrium transport. How such exchange induced spin polarization is retained, or enhanced, by an absence of momentum degeneracy is puzzling. However, for practical applications, it is now possible to vary the degree of spin polarization in a way not previously possible. A complex arrangement of gates and applied voltages can be utilized for on-chip spin manipulation with applications in spintronics and quantum information processing.

%%%%%%%%%%%%%%%%%%%%%%%%%%%%%%%%%%%%%%%%%%%%%%%%%%%%%%%%%%%%%%%%%%%%%
%% The "Acknowledgement" section can be given in all manuscript
%% classes.  This should be given within the "acknowledgement"
%% environment, which will make the correct section or running title.
%%%%%%%%%%%%%%%%%%%%%%%%%%%%%%%%%%%%%%%%%%%%%%%%%%%%%%%%%%%%%%%%%%%%%

\section*{Acknowledgements}
We thank L.W. Smith, K.-F. Berggren, and C.-T. Liang for useful discussions. For technical assistance, we thank D. Anderson and S.-J. Ho. This work was supported by the Engineering and Physical Sciences Research Council (UK), EU Spinmet, and the National Science Council (Taiwan). I.F. acknowledges support from Toshiba Research Europe.


\begin{thebibliography}{99}
\bibitem{note1} Differential conductance of the collector was measured at a constant current.This prevents the one-dimensional channel from being fully pinched off. When the dc bias current is high ($I_{sd}=30$~nA)  the lowest conductance is $0.25(2e^2/h)$ and consequently a 0.25 plateau is not observed although the spin polarization is identical. Also note that at a fixed dc current bias, $I_{sd}$, the corresponding dc voltage bias varies with the conductance.
\bibitem{note2} The collector differential conductance at a particular gate voltage is dependent on source-drain bias and, therefore, normalization is not possible for different $I_{sd}$.
\end{thebibliography}
\end{document}